\documentclass[aps,pre,onecolumn]{revtex4}

\usepackage{graphicx}
\usepackage{color}
\usepackage{graphicx}
\usepackage{amsmath}
\usepackage{dcolumn}

\usepackage{epsfig}
\usepackage{graphics}
\usepackage{latexsym}
\usepackage{amsfonts}
\usepackage{amssymb}

\newcommand{\be}{\begin{eqnarray}}
\newcommand{\ee}{\end{eqnarray}}

\newcommand{\fra}[1]{ { \color{black} #1}} 

\newcommand{\el}{{\rm  e}}
\newcommand{\ph}{{\rm  p}}
\newcommand{\phs}{{\rm s}}

\newcommand{\css}{{\rm c}}
\newcommand{\dss}{{\rm d}}
\newcommand{\rss}{}

\newcommand{\bc}{\sigma_{\ph}}
\newcommand{\bce}{\sigma_{\el}}
\newcommand{\bci}{\sigma_{i}}

\newcommand{\Tti}{\tilde{T}}

\newcommand{\Te}{T_{\el}}
\newcommand{\Tp}{T_{\ph}}
\newcommand{\Ts}{T_{\phs}}
\newcommand{\Ttip}{\Tti_{\ph}}
\newcommand{\ke}{k_{\el}}
\newcommand{\kp}{k_{\ph}}
\newcommand{\ks}{k_{\phs}}
\newcommand{\ale}{\alpha_{\el}}
\newcommand{\alp}{\alpha_{\ph}}
\newcommand{\als}{\alpha_{\phs}}
\newcommand{\ce}{c_{\el}}
\newcommand{\cp}{c_{\ph}}
\newcommand{\cs}{c_{\phs}}
\newcommand{\Me}{M_{\el}}
\newcommand{\Mp}{M_{\ph}}
\newcommand{\bip}{B_{\ph}}
\newcommand{\bie}{B_{\el}}

\newcommand{\tauce}{\tau_{\css \el}}
\newcommand{\taucp}{\tau_{\css \ph}}
\newcommand{\tauci}{\tau_{\css i}}
\newcommand{\taude}{\tau_{\dss \el}}
\newcommand{\taudp}{\tau_{\dss \ph}}
\newcommand{\taudi}{\tau_{\dss  i}}
\newcommand{\taure}{\tau_{\rss \el}}
\newcommand{\taurp}{\tau_{\rss \ph}}
\newcommand{\tauri}{\tau_{\rss  i}}

\newcommand{\tcar}{\tau}
\newcommand{\tcarmin}{ \tau_{{\rm min}}  }
\newcommand{\tcarmax}{ \tau_{{\rm max}}  }

\newcommand{\mat }{\mathsf{M}}
\newcommand{\vecT}{\mathsf{T}}
\newcommand{\trace}{\mathrm{Tr}}
\newcommand{\determ}{\mathrm{D}}

\newcommand{\bcunit}{W\,m$^{-2}$\,K$^{-1}$}
\newcommand{\bcunitM}{MW\,m$^{-2}$\,K$^{-1}$}
\newcommand{\tauh}{\tau_{1/2}}

\newcommand{\teff}{\tau_{{\rm eff}}}

\newcommand{\itf}{-}

\begin{document}

\title{Influence of the electron-phonon interfacial conductance on the thermal transport at metal/dielectric interfaces.}

\author{J. Lombard, F. Detcheverry and S. Merabia}
\affiliation{Institut Lumi\`ere Mati\`ere, UMR5306 Universit\'e  Lyon 1-CNRS, Universit\'e de Lyon 69622 Villeurbanne, France}
\date{\today}

\begin{abstract}
Thermal boundary conductance  at a metal-dieletric interface
is a quantity of prime importance for heat management at the nanoscale. 
While the boundary conductance is usually ascribed to the coupling between metal phonons and dielectric phonons, 
in this work we examine the influence of a direct coupling between the metal electrons and the dielectric phonons.
The effect of electron-phonon processes is generally believed to be resistive, and tends to decrease the overall 
thermal boundary conductance as compared to the phonon-phonon conductance $\bc$. Here, we find that
the effect of a direct coupling $\bce$ is to enhance the effective thermal conductance, between the metal and the dielectric.
Resistive effects turn out to be important only for thin films of metals having a low electron-phonon coupling strength.
Two approaches are explored to reach these conclusions. 
First, we present an analytical solution of the two-temperature model
to compute the effective conductance which account for all the relevant energy channels, 
as a function of $\bce$, $\bc$ and the electron-phonon coupling factor $G$. 
Second, we use numerical resolution to examine the influence of $\bce$ on two realistic cases:
gold film on silicon or silica substrates. 
We point out the implications for the interpretation of time-resolved thermoreflectance experiments. 

\end{abstract}

\maketitle

\section{Introduction}
Interfacial thermal transport is of major importance in heat application management. 
Indeed, ohmic contacts are ubiquitous in microelectronics 
and experience increasing levels of loads. 
As a result, a fundamental understanding of the cooling kinetics of heated metals 
or semiconductor at ohmic contacts 
is a prerequisite to design electronic components that can support large heat flux densities~\cite{pop2010}.  
Thermal relaxation of microelectronics components is primarily governed by the boundary conductance, 
since as the density of interfaces increases, thermal losses in the bulk become negligible 
compared to interfacial resistance. 
While the boundary resistance involves {\em a priori} the energy transfer between metal electrons or metal phonons,
and the phonons in the substrate, 
most of the models developed so far consider only the latter channel of energy. 
These include the acoustic mismatch model (AMM), 
where phonon transmission at an interface is assumed to be controlled by 
the difference in the acoustic impedances of the two media~\cite{little1959}, 
and the diffuse mismatch model (DMM) which posits that interfacial roughness 
destroys any correlation between incident and reflected phonons~\cite{swartz1989}.

The boundary resistance of metal-diamond interfaces at room temperature has been first experimentally determined by Stoner and Maris in 1993~\cite{stoner1993}, using 
time-resolved thermoreflectance (TTR). This work has been followed by the characterization of various interfaces, 
extending the early measurements to higher temperatures~\cite{lyeo2006,komarov2003,stevens2005,monachon2014}. More recently, TTR has been applied to investigate heat transfer across metal-graphite~\cite{schmidt2010} and metal-carbon nanotube interfaces as well~\cite{hopkins2012}. In parallel, theoretical progress has been achieved through molecular dynamics~\cite{termentzidis2011,dacruz2012} which ignores electronic degrees of freedom, or treat them in the spirit of the two-temperature model~\cite{wang2012b}. A recurrent conclusion of this body of works is that the Kapitza conductances measured between two solids
depart significantly from the classical AMM and DMM models, 
which often yield comparable values.   
For instance, the thermal conductance of one of the most studied system, the Pb\itf diamond interface,    
is found to be typically sevenfold higher than the DMM prediction.

\begin{figure}[htbp]
\includegraphics[scale=0.4]{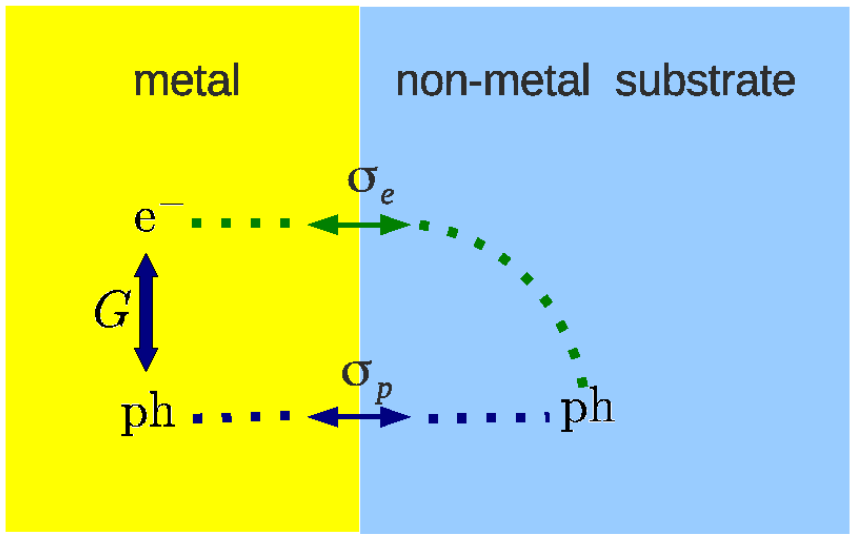}
\caption{(Color online) Schematic representation of the two energy transfer channels between metal electrons (e$^{-}$) and substrate 
phonons (ph). The indirect coupling (blue) involves electron-phonon coupling in the bulk metal, characterized by $G$ followed 
by transfer via the phonon-phonon conductance $\bc$. This channel has been studied by Majumdar and Reddy~\cite{majumdar2004}. The direct route studied in this work and represented in green, involves the electron-phonon conductance $\bce$, for which theoretical predictions exist~\cite{huberman1994,sergeev1998,mahan2009}. Diffusive transport represented by dots plays a role in both cases.} 
\label{fig_different_channels}
\end{figure}

After the early room temperatures measurements of Stoner and Maris, 
several authors have proposed to relate this excess of conductance 
to electron-phonon scattering taking place across the interface~\cite{huberman1994,sergeev1998,majumdar2004,mahan2009}. 
Indeed, there are at least two mechanisms according to which electrons may heat up the substrate, as represented 
in Fig.~\ref{fig_different_channels}: The first one has been extensively studied and involves an indirect coupling between 
the electrons and the substrate through phonon-phonon processes~\cite{majumdar2004,jones2013,hopkins2013,wang2012b,singh2013,wilson2013,dechaumphai2014}. This mechanism is only operative when the electrons are not in equilibrium with the lattice. In this situation, the effect of the electron-phonon process is to introduce an additional resistance $1/\sqrt{G \kp}$ where $G$ is the electron-phonon coupling factor, and $\kp$ the phononic thermal conductivity of the metal. Hence, it is found that electron-phonon processes deteriorate the interfacial energy transfer, as compared to the case where electronic degrees of freedom are neglected.  
%
%

There is another energy channel that has been recently unveiled from the experimental side~\cite{hopkins2009,guo2012,li2012}, and which has been previously predicted theoretically~\cite{huberman1994,sergeev1998,mahan2009}. This mechanism is a direct coupling of the electron-phonon coupling through the interface, characterized by an interfacial heat flux of the form $\bce (\Te-\Ts)$, 
which depends on the temperature difference between the electrons and the phonons substrate, through the electron-phonon interfacial conductance $\bce$.  Throughout the  decade $2000$ and until recently, 
the latter electron\itf phonon transfer mechanism has been disregarded essentially
because experiments where the metal is replaced by Bi a semi-metal and which conclude that the values of the 
interfacial conductance seems to be independent on the nature of the electronic transport properties of the film~\cite{lyeo2006}.  
There are, however, actually two hints that the {\em direct} electron-phonon channel could play a significant role. 
First, the above mentioned models for the conductance $\bce$ 
predict values that are comparable or even largely exceed the typical values for the phonon-phonon conductance $\bc$.  
Second, recent transient thermoreflectance experiments on a Au film~\cite{hopkins2007,hopkins2009,guo2012}
concluded on the existence of an electron-substrate energy transfer, 
with  estimate of $\bce$ in the range $100$ to $1000$ MW m$^{-2}$ K$^{-1}$, compatible with the theoretical predictions.
The consequences of this direct 
channel of energy have not been discussed in the literature. 

In this article, we aim at a theoretical description of the competition between the three different channels represented in Fig.~\ref{fig_different_channels}. We show that electron-phonon processes taking place in the bulk of the metal induce
a small, commonly negligible resistance, while the effect of the direct coupling is to enhance interfacial heat transfer. Importantly, we find that the effect of a direct coupling persists over long time scales on the nanosecond range, even though the electrons and the phonons in the metal have been long allowed 
to equilibrate. Indeed, if the largest values of $\bce$ are experimentally reported in situations where electrons are out of equilibrium 
with the metal lattice~\cite{guo2012}, still a non vanishing value $\bce > 100$ MW m$^{-2}$ K$^{-1}$ is measured in case of negligible electron-phonon nonequilibrium in the metal~\cite{hopkins2009}.
This indicates that this channel of interfacial transfer should be operative in the TTR measurements of metal\itf dielectric interfaces, and may explain the large discrepancies observed between the DMM predictions and the TTR measurements.

The article is organized as follows:
In Sec.~\ref{sec:motivations}, 
we first motivate the study by calculating the interfacial electron-phonon conductance for different interfaces using Sergeev's model
for $\bce$. 
%
Using analytical calculations on the two-temperature model,    
we examine in Sec.~\ref{sec:anal} the influence of a finite $\bce$ on the temperature decay of a heated metal. 
In Sec.~\ref{sec:realistic}, we illustrate the effect of $\bce$ for two specific systems: 
gold/silicon and gold/silica interfaces.
We show that for electron\itf phonon conductance larger than the phonon\itf phonon conductance, 
it is actually the dominant mechanism governing the temperature decay. 
Our conclusions are given in Sec.~\ref{sec:con}. 

\section{Theoretical estimates for electron-phonon and phonon-phonon conductances}
\label{sec:motivations}

In this section, we motivate our study by estimating, for various metal/non-metal systems, 
the  electron-phonon and phonon-phonon interfacial conductances. 
In both cases, we rely on theoretical predictions.

\paragraph*{Electron-phonon conductance $\bce$. }
There are three analytical models predicting  the electron-phonon interfacial conductance at a metal/dielectric interface. 
Huberman and Overhauser~\cite{huberman1994} first propose that metal electrons couple 
with joint vibrational modes at the interface leading to a significant enhancement of the energy transfer, 
as compared with direct phonon-phonon  coupling.  
Later, Sergeev developed a field theory to predict the effect of the boundaries on the inelastic scattering of conduction electrons~\cite{sergeev1998}.
Finally, Mahan derived a theory where metal electrons may transfer heat to the substrate 
through the image charges created by the vibrating ions of the non-metal~\cite{mahan2009}. 
This model should mainly apply to ionic crystals or polar semi-conductors, 
rather than to apolar non-metals. 
In both Huberman and Mahan's models
the conductance is predicted to be  constant at high temperatures, 
while in Sergeev's model,
it is found to increase linearly with 
temperature~\footnote{Such temperature dependence stems from the inelastic nature of electron scattering at the interface.}, 
in agreement with experimental observations~\cite{hopkins2009}. 
Because it is the sole model to capture this temperature dependence, 
we chose to consider only Sergeev's model in the following. 

As far as we know, there has been no attempt to quantify the values of the electron-phonon interfacial conductance 
predicted by Sergeev's model for real cases at ambient temperatures. 
Here we  determine the values of $\bce$ for different metal/dielectric interfaces. 
We do so by using input data from density function theory (DFT) calculations predicting the electron-phonon coupling of several noble metals. 
To proceed, we start with the expression
\begin{equation}
\label{sigma_sergeev}
\bce = \frac{3 \pi \hbar}{35 \zeta(3) k_B} \frac{\gamma u_L}{\tau_{\rm e-p}}\left[  1+2\left(\frac{u_L}{u_T}\right)^3 \right],
\end{equation}
where  $\hbar$ is the reduced Planck constant, 
$k_B$ is Boltzmann constant, 
$\zeta$ is Riemann zeta function, 
$\gamma$ is the Sommerfeld constant of the metal,  
$u_L$ and $u_T$ are respectively the metal longitudinal and transverse sound velocities 
and $\tau_{\rm e-p}$ is the electron-phonon energy relaxation time of the bulk metal. 
Because at high temperature $1/\tau_{\rm e-p} \sim T_\el$~\cite{lin2008}
while other factors have weak or no temperature dependence, 
one immediately obtains $\bce \sim T$.  
%
Here, we estimate the electron-phonon energy relaxation time using its relation with the electron-phonon coupling factor~\cite{lin2008,kaganov1957}:
\begin{equation}
G=\frac{\pi^2 m_\el \, c_s^2 \, n_\el}{6 \, \tau_{\rm e-p} T_\el},
\end{equation}
where $m_\el$ is the effective mass of the electron, $c_s$ is the speed of sound and $n_\el$ is the free electrons number density.
Besides, one has:
\begin{equation}
\gamma_T \equiv \frac{G}{C_\el}=\frac{3 \hbar \lambda \langle \omega^2 \rangle}{\pi k_B T_\el},
\end{equation}
where $\lambda$ is the electron-phonon coupling constant,  
so that the electron-phonon conductance Eq.~\ref{sigma_sergeev} may be cast as
\begin{equation}
\label{sigma_sergeev2}
\bce = \frac{54 \gamma^2 \hbar^2 \lambda \langle \omega^2 \rangle}{35 \pi^2 k_B^2 m_\el c_s n_\el} T_\el \left[  1+2\left(\frac{u_L}{u_T}\right)^3 \right].
\end{equation} 
To calculate $\bce$, we have used the values of $\lambda \langle \omega^2 \rangle$ 
from the DFT calculations of Lin et {\em al.}~\cite{lin2008} for noble metals. 
The effective mass $m_{\rm e}^{\star}$ is taken equal to  the electron  mass $m_{\rm e}$, 
except for Platinum, for which we have used $m_{\rm e}^{\star}=13$~$m_{\rm e}$~\cite{Kasap}. 
For Bismuth, we have calculated the value of $\lambda \langle \omega^2 \rangle$ from the 
thermodynamic analysis of Giret et {\em al.}~\cite{giret2011,gamali2013}, 
who reported an electron-phonon coupling factor $G=1.45$ $10^{16}$ W m$^{-2}$ K$^{-1}$. 
Table~\ref{tab:metal_properties} gathers the values of the electron-phonon interfacial conductance 
for common metals and the semi-metal Bi. 


\begin{table}[ht]
\begin{tabular}{c r r r c c c c r }
\hline
\hline
metal  & $n$\ \ \ \ \    & $u_L$  & $u_T$  & $n_\el$ &  $\gamma$ & $\lambda \langle \omega^2 \rangle $ & $\bce$  \\ 
\hline
Al  & 100000  & 6240 & 3040 & 18.1 & 1.3   & 185.9 & 2.64\\              
Au  & 97970   & 3390 & 1290 & 5.9  & 0.67  & 23 $\pm$ 4 & 3.08 \\
Bi  & 46794   & 1543 & 1107 & 14.1 & 0.08 & 229.3 & 0.02 \\
Cr  & 138269  & 6980 & 4100 & 16.6 & 1.57  & 128 & 6.2  \\
Pb  & 55990   & 2350 & 970  & 13.2 & 2.93  & 45 $\pm$ 5 & 17.8\\
Pt  & 110872  & 4174 & 1750 & 3.2  & 6.54 & 142.5 & 186 \\
\hline 
\hline
\end{tabular}
\caption{Physical parameters for the metals: 
molar density $n$ (mol\,m$^{-3}$), 
longitudinal and transverse sound velocities $u_L$ and  $u_T$ (m\,s$^{-1}$), 
free electron number density $n_\el$ (10$^{28}$ m$^{-3}$) , 
Sommerfeld's constant $\gamma$ (mJ\,mol$^{-1}$\,K$^{-2})$, 
factor $\lambda \langle \omega^2 \rangle$ (meV$^2$), 
and electron-phonon conductance $\bce$  (100  \bcunitM) at room temperature. 
}
\label{tab:metal_properties}
\end{table}

\paragraph*{Phonon-phonon conductance $\bc$.}
Here we  use the diffuse mismatch model~\cite{swartz1989}, as we briefly discuss now. 
Quite generally, the phonon-phonon thermal boundary conductance is given by
\begin{widetext}
\begin{equation}
\label{general_conductance}
\bc=\frac{1}{2} \sum_p \int_{0}^{\omega_{1,p}^c} v_{1,p} \bar {h} \omega D_{1,p}(\omega) \frac{\partial f_{\rm eq}(\omega,T)}{\partial T} \int_{0}^{\pi/2} \alpha_p(\omega,\cos \theta) \cos \theta d (\cos \theta) d \omega, 
\end{equation}
\end{widetext}
where the sum runs over the polarizations $p$, $\omega_{1,p}^c$ is the cut-off frequency, $v_{1,p}$ is the mode group velocity, 
$D_{1,p}(\omega)$ is the density of states, 
$f_{\rm eq}$ is the equilibrium Bose-Einstein distribution, 
and $\alpha_p(\omega,\cos \theta)$ is the phonon transmission coefficient. 
If interfacial scattering is supposed to be diffuse, the transmission coefficient 
no longer depends on the angle of incidence, and from detailed balance it follows that
\begin{equation}
\label{transmission_coefficient}
\alpha=\frac{\sum_p \frac{1}{v_{2,p}^2}}{\sum_p \frac{1}{v_{1,p}^2} + \sum_p \frac{1}{v_{2,p}^2}}, 
\end{equation}
where the density of states has been taken to be given by Debye's model. 
Note that the transmission coefficient is assumed to be polarization independent~\cite{duda2010}. 
To compute $\bc$, we have used the acoustic properties of the different metals and dielectric 
provided in Tabs.~\ref{tab:metal_properties} and~\ref{tab:substrate_properties}.
 
\begin{table}[h]
\begin{tabular}{ c c c c c c c c c }
\hline
\hline 
substrate & $n$ (mol\,m$^{-3}$)  & $u_L$ (m\,s$^{-1}$)  & $u_T$ (m\,s$^{-1}$)    \\ 
\hline
Al$_2$O$_3$ & 38921   & 10890 & 6450 \\
AlN         & 160345  & 11120 & 6267 \\
diamond     & 292667  & 17500 & 12800\\
GaN         & 138269  & 7781  & 4427 \\
Si          & 73214   & 8970  & 5332 \\
SiO$_2$     & 44167   & 5950  & 3740 \\
\hline 
\hline
\end{tabular}
\caption{Physical parameters of the substrates: 
molar density $n$, longitudinal and transverse sound velocities $u_L$ and  $u_T$.} 
\label{tab:substrate_properties}
\end{table}

\begin{figure*}[htbp]
\includegraphics[scale=0.4]{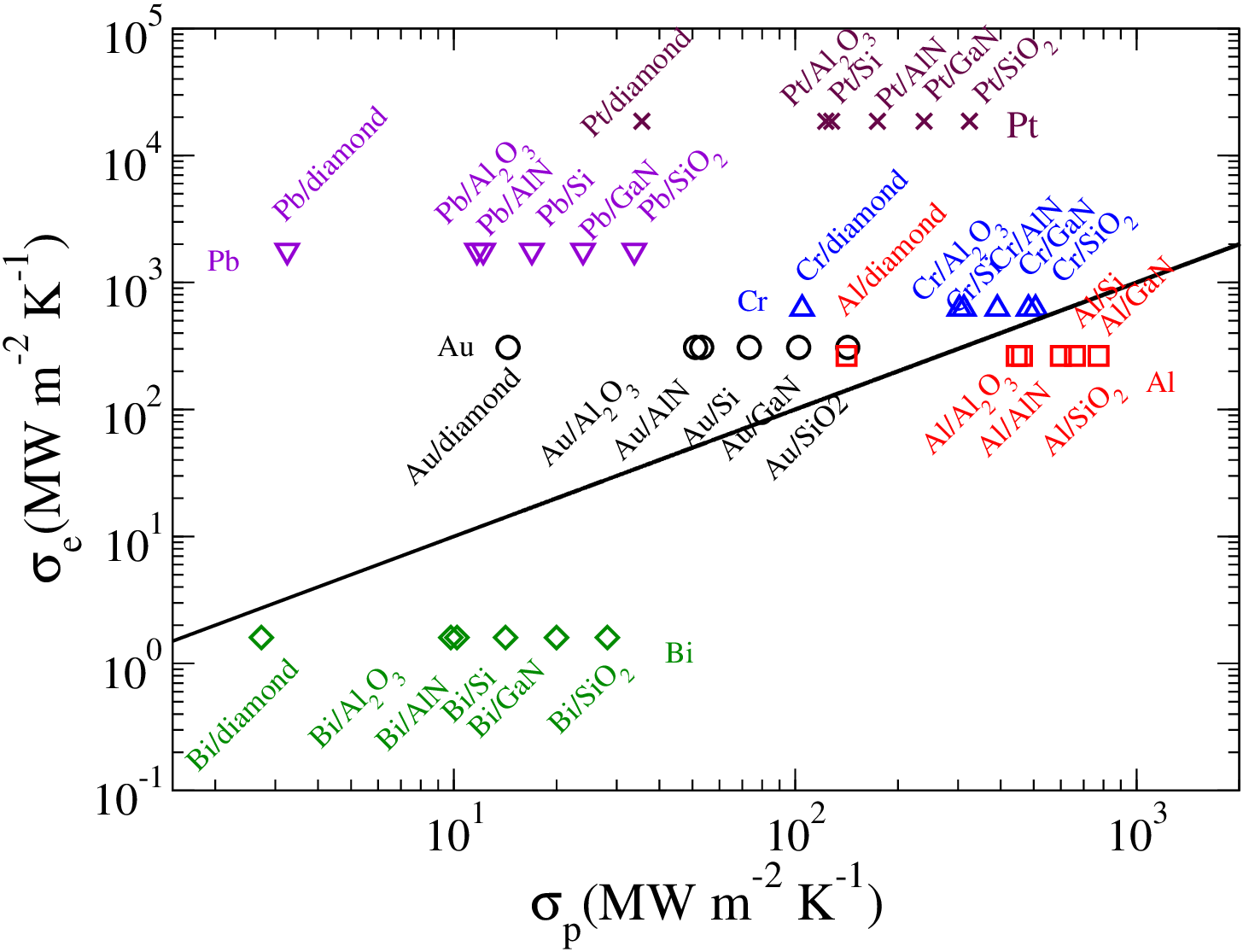}
\caption{(Color online) Phonon-phonon conductance $\bc$ and electron-phonon conductance $\bce $ 
estimated from  Sergeev~\cite{sergeev1998} and the DMM model respectively, at room temperature.  
The black solid line indicates the curve $\bce=\bc$.} 
\label{fig_diagram_hep_hpp}
\end{figure*}

Figure~\ref{fig_diagram_hep_hpp} gathers the electron-phonon and phonon-phonon conductances for the different systems considered.  
It is apparent that in most cases $\bce$ is much larger than $\bc$.  
This implies that for most of the metal/dielectric interfaces, $\bce$ may not be neglected, 
and should be accounted for in the analysis of thermoreflectance data. 
Qualitatively, one expects that  compared to the bare phonon-phonon conductance, 
the apparent thermal conductance of the systems discussed should be enhanced, 
as will be investigated quantitatively in the next sections. 
Besides, note that even Bismuth which is a semi-metal, and not considered a good metal 
is predicted to have a significant $\bce$. 
This suggests that the relevant energy channel discussed here should be present not only in metals, but also in semi-metals
despite the fact that the number of free electrons is relatively small. 
Hence, even materials which display poor electronic transport properties 
such as semi-metals, could be affected by electron-phonon interfacial coupling. 
This is because for these materials the electron-phonon relaxation time takes values slighty larger but comparable with those of good metals. 

It should be kept in mind however, that our estimates of the transport coefficient $\bce$ comes from a model which may have some limitations.
In particular, Sergeev's prediction yields a conductance which depends only on the bulk properties of the metal, 
and not on the nature of the substrate.
Also, it should be remarked that for gold systems, 
Sergeev's model overestimates by a factor of two 
the electron-phonon conductance reported experimentally, as shown below in Fig.~\ref{fig_electron_phonon_conductance}. 
Further work is clearly needed to develop models of interfacial electron-phonon coupling, which includes information from the interface. 
\fra{However, even if Sergeev's model may overestimate $\bce$, 
$\bce$ remains comparable to $\bc$ in many cases, 
suggesting that from those theoretical estimates, 
there is no reason to discard the electron-phonon conductance. 
Given the uncertainty of theoretical estimates for $\bce$, 
and the lack of experimental measurements for most systems, 
we will not choose  any specific value or model, 
but rather explore a wide range of values for $\bce$.}


\begin{figure}[h]
\includegraphics[scale=0.31]{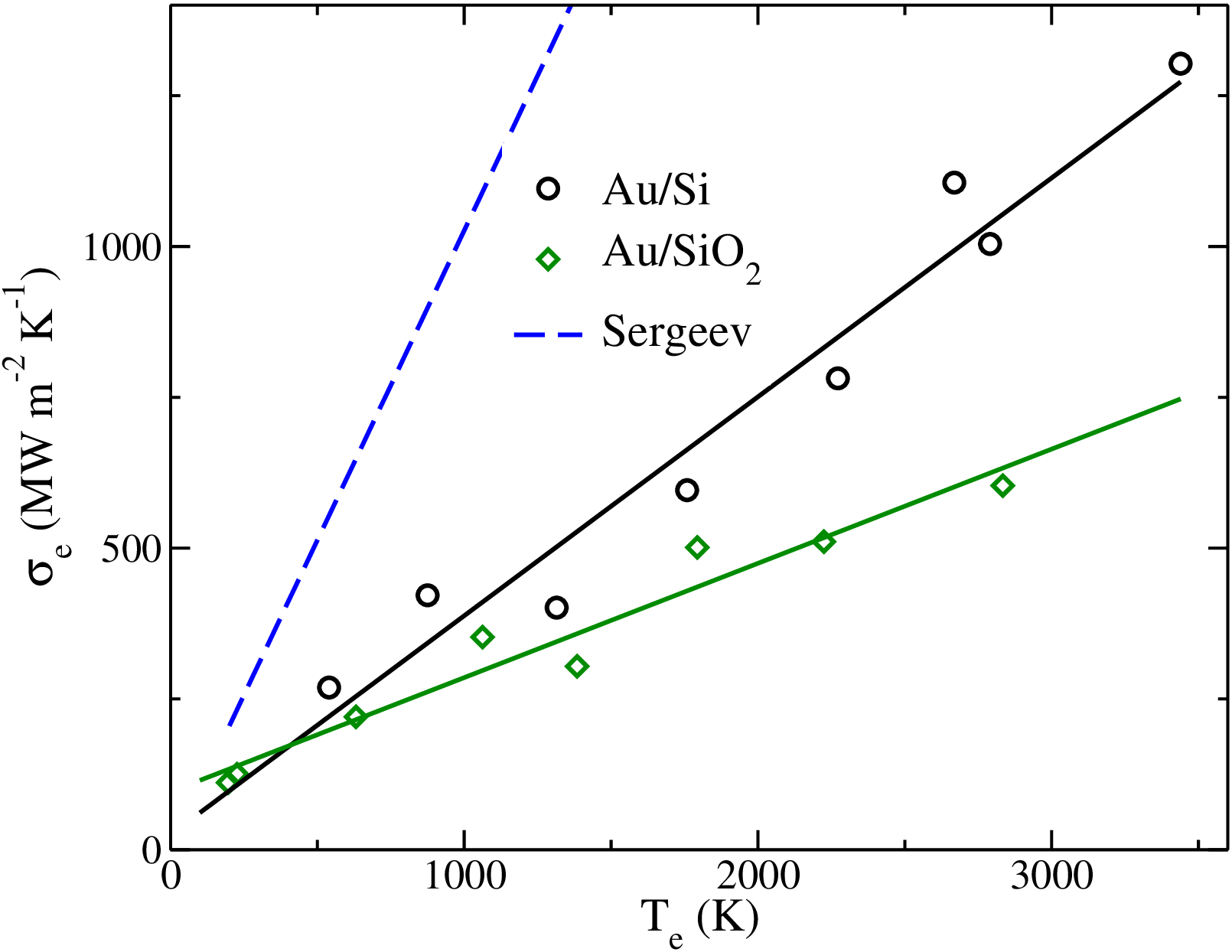}
\caption{(Color online) Temperature dependence of the electron-phonon interface conductance 
for gold/silica and gold/silicon, as reported experimentally~\cite{hopkins2009}. 
The solid lines are the values obtained from Eqs.~\ref{bcpexpsilicon} and~\ref{bcpexpsilica}.
The blue dashed line shows Sergeev's predictions for gold, as given by eq.~\ref{sigma_sergeev2}.}
\label{fig_electron_phonon_conductance}
\end{figure}

\section{Analytical predictions}
\label{sec:anal}

We derive in this section analytical expressions for the characteristic cooling time of a heated metal/dielectric interface, 
taking into account not only the phonon-phonon conductance, 
but also the electron-phonon conductance. 
The analytical solutions may allow to understand quantitatively the effect of $\bce$ 
on the effective interfacial conductance which account for the different channels, and analyze the limiting behaviours.

\subsection{Model and parameters}
\begin{figure}
\includegraphics[width=0.5\linewidth]{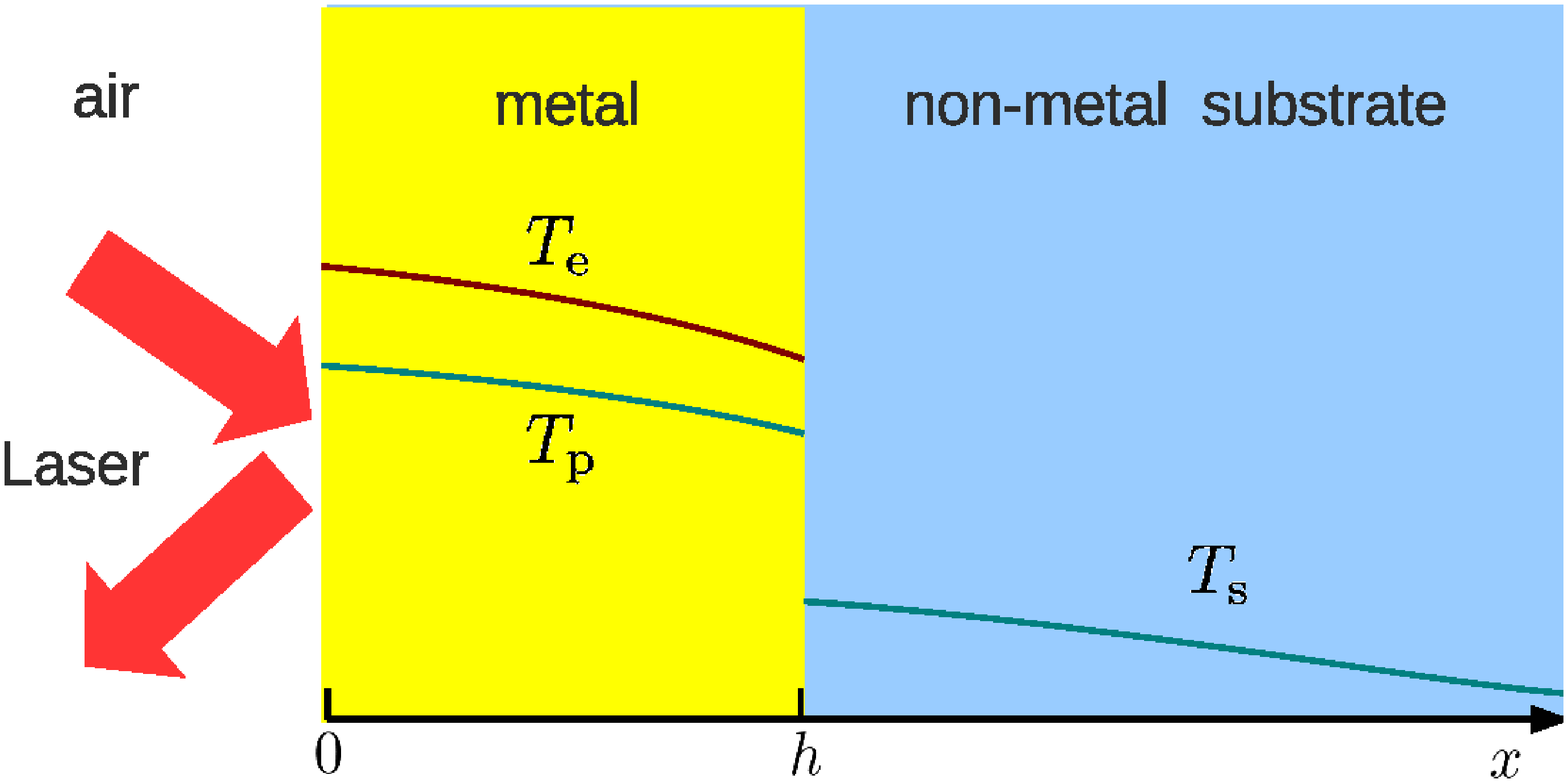}
\caption{(Color online) Sketch of the system modeled: 
a thin metallic film on a dielectric substrate, 
and irradiated by an ultrafast laser pulse. 
$\Te$, $\Tp$  are the  electron and phonon temperatures in the metal, 
$\Ts$ is the phonon temperature in the substrate. 
}
\label{fig:schematic}
\end{figure}

The situation relevant to TTR experiments is schematically depicted in Fig.~\ref{fig:schematic}: 
a metal film with thickness $h$ on a  semi-infinite dielectric 
substrate~\footnote{The dielectric substrate is not infinite, 
but has finite length $H$ in the $x$-direction. 
However, the associated time scale $H^2/\als$ is so large that it is generally irrelevant in experiments.}.
To keep calculations tractable, 
we assume that the problem is one-dimensional, with variation occurring only perpendicular to the interface ($x$-direction). 
In principle, the cooling kinetics should be described by the solution of the Boltzmann transport equation, but the long time regime on the order of $1000$ ps that we discuss here is generally well captured by the two-temperature model where the electrons and phonons are each characterized by a temperature, subject to diffusion and relaxation~\cite{kaganov1957,allen1987,wilson2013}. 
The evolution equations are
\be
\ce \, \partial_t \Te &=&  \ke \, \partial_{xx}^2 \Te \, - \, G(\Te - \Tp), \nonumber  \\
\cp \, \partial_t \Tp &=&  \kp \, \partial_{xx}^2 \Tp \, + \, G(\Te - \Tp), \nonumber  \\
\cs \, \partial_t \Ts &=&  \ks \, \partial_{xx}^2 \Ts.                      \nonumber
\ee
Here, we use subscripts $i=\el$, $\ph$, and $\phs$ for the metal electrons, 
the metal phonons and the substrate phonons respectively, 
$c_i$ and $k_i$ are the heat capacities and thermal conductivities, 
$G$ is the coupling factor. 
Unless otherwise mentioned, we will consider that all coefficients are constant, and  in particular independent of temperature.  
All quantities are given in SI units, unless indicated otherwise.
As regards the boundary conditions, we assume a zero energy flux at the air\itf metal interface, 
while the metal\itf dieletric interface is characterized by {\it two} boundary conductances,
\be
\mbox{for\ }x=0,  \quad  -\ke \, \partial_x \Te  &=& -\kp \, \partial_x \Tp =0,                              \nonumber \\ 
                                       &&                                                          \nonumber \\ 
\mbox{for\ }x=h,  \quad  -\ks \, \partial_x \Ts  &=& \bce (\Te-\Ts) + \bc (\Tp-\Ts),                        \nonumber \\ 
                                 &=& -\ke \,  \partial_x \Te \: - \: \kp  \,  \partial_x \Tp.       \nonumber
\ee
Here $\bc$ and $\bce$ are the phonon-phonon and electron-phonon conductances respectively 
and the second equation expresses the continuity of the energy flux across the interface. 
Finally,  the initial conditions are 
\be
\mbox{for\ } t=0,  \quad \Te= T^i(x) ,  \quad \Tp=\Ts=0.
\ee
$\Te$, $\Tp$ and $\Ts$ are thus the differences with respect to 
the temperature $T_0$ of the system prior to any laser heating~\footnote{Because all coefficients are assumed to be constant, $T_0$ is irrelevant.}.  
The initial temperature profile $T^i$ is exponential with decay length $\delta$,  the penetration depth of the laser in the metal film: 
\be
T^i(x)  &=& T^i(0)\: e^{-x/\delta},  \qquad   T^i(0)= \frac{h}{\delta (1 -e^{-h/\delta})}.
\nonumber
\ee
The normalization constant $T^i(0)$ is such that the average temperature increase in the film is 
unity~\footnote{The normalisation constant is irrelevant when the long-time decay is exponential but will play a role when the decay is power-law.}. 
Below, we focus on the electron temperature at the surface of the film, 
and its long-time behavior~\footnote{In experiments, what is probed is actually a combination of electron and phonon temperatures.}.  

\begin{table*}[htb]
  \begin{tabular}{l l l l l }
     \hline \hline
     Quantity         &    Symbol              & Definition     &   Value    &   Unit  \\   
     \hline
     film thickness   & $h$                    &   -             &  100                                        &   nm   \\ 
     penetration depth& $\delta$               &   -             &   20                                        &   nm   \\ 
     heat capacities  & $\ce$, $\cp$, $\cs$    &   -             & $10^4$, $10^6$, $10^6$                      &   J\,m$^{-3}$\,K$^{-1}$  \\  
     thermal conductivities   & $\ke$, $\kp$, $\ks$    &   -             & $10^2$, 10, $10^2$                  &   W\,m$^{-1}$\,K$^{-1}$  \\   
     coupling factor  & $G$                    &   -             & $10^{16}$                                   &   W\,m$^{-3}$\,K$^{-1}$  \\  
     conductances     & $\bce$, $\bc$          &   -             & $10^{5-11}$, $10^{7-9}$                     &   \bcunit  \\  
     \hline
     diffusivities    & $\ale$, $\alp$, $\als$\ \ \ &  $\alpha_i=k_i/c_i$      & $10^{-2}$, $10^{-5}$, $10^{-4}$     &   m$^2$\,s$^{-1}$ \\    
     conductance times\ \ \ \ & $\tauce$, $\taucp$   &  $\tauci= c_i h/\bci$    & $10^{4}-10^{-2}$, $10^{4}-10^{2}$   &  ps \\
     diffusion  times  & $\taude$, $\taudp$          &  $\taudi=h^2/\alpha_i $  & 1, 1000                             &  ps \\
     relaxation times  & $\taure$, $\taurp$          &  $\tauri=c_i/G$          & 1, 100                              &  ps \\ 
     \hline       
     Biot numbers     &  $\bie$, $\bip$ &  $B_i=\taudi/\tauci= h \bci/k_i$  \ \ \  & $10^{-4}-10^{2}$, $0.1-10$  & -  \\
     M numbers        &  $\Me$, $\Mp$   &  $M_i=\taudi/\tauri=G h/\bci $           & 1, 10                       & -  \\
     R number         &  $R$            &  $R=\taurp/\taure=\cp/\ce$               & 1000                        & -  \\
   \hline\hline
  \end{tabular}
\caption{Quantities used in this work. 
Subscripts i=$\el$, $\ph$, or $\phs$ refer to the metal electrons, 
the metal phonons and the substrate phonons respectively.
}
\label{tab:cases}
\end{table*}

Table~\ref{tab:cases} lists the various quantities needed  in the problem, 
from which one can construct the relaxation, diffusion and conductance times. 
The relaxation time $\tauri=c_i/G$ is related  to the electron-phonon coupling
in the bulk of the metal film,  
$\taudi=h^2/\alpha_i$ is the typical time for diffusion over the film, 
and $\tauci= c_i h/\bci$ characterizes  the cooling of the 
carriers induced by the interfacial conductances. 
Taking ratios between those characteristic times  leads to dimensionless parameters:
the $R$ number compares the electron and phonon relaxation times,
the Biot numbers $B_{i=\el,\ph}$ compare the diffusion and conductance times  
and the $M_{i=\el,\ph}$ numbers  compare  the relaxation and diffusion times. 
Below we focus on the influence of the two interface conductances, 
with $\bce$ taken in the broad range $10^5-10^{11}$, 
and $\bc$ in the range $10^7-10^9$ \bcunit, 
which should encompass most cases of experimental relevance.
All other parameters are kept fixed to a value representative of real systems, 
those default values are also given in Tab.~\ref{tab:cases}.

\subsection{Influence of the electron-phonon conductance $\bce$}
We now investigate how, within the two-temperature model, 
a finite electron-phonon conductance  $\bce$ affects the temperature decay. 
We first present a simplified model, whose analytical solution is straightforward.

\subsubsection{Fast diffusion approximation }
To obtain compact formulas, we consider here the two-temperature model in the fast diffusion (FD) approximation:  
the temperature is assumed to be uniform within the film, implying that both electrons and phonons diffusion are neglected. 
With the vector notation $\vecT=[\Te, \Tp]^T$, where $T$ denotes the transpose, the equations are 
\be
\partial_t \vecT = - \mat  \vecT, \qquad
\mat = 
 \begin{bmatrix}
  \frac{1}{\taure} + \frac{1}{\tauce}   &  -\frac{1}{\taure}                    \\
  -\frac{1}{\taurp}                     &  \frac{1}{\taurp} + \frac{1}{\taucp}  \\
 \end{bmatrix} 
. \nonumber
\ee
The decay is found to be exponential, with a characteristic decay time $\tcar$ given by 
\be
\tcar   &=& \frac{2}{\trace - \sqrt{\trace^2 - 4 \determ} },  \label{eq:tcartoy}  \\
\trace  &=& {\rm Tr} (\mat) = \frac{1}{\taure} + \frac{1}{\taurp} + \frac{1}{\tauce} +  \frac{1}{\taucp}, \nonumber \\
\determ &=& {\rm Det} (\mat) = \frac{1}{\taure \taucp} + \frac{1}{\taurp \tauce} + \frac{1}{\tauce \taucp},   \nonumber 
\ee
To get a qualitative understanding of the $\tcar$ dependence on $\bce$, 
we first consider the two limiting values for  
$\bce \rightarrow 0$ and 
$\bce \rightarrow \infty$, 
which are denoted as $\tcarmax$ and $\tcarmin$. 
Physically, $\tcarmax$ represents the metal's cooling time 
in the absence of any electron-phonon conductance, 
while $\tcarmin$ is the cooling time in the limit of high electron-phonon conductance. 
On physical grounds,  one can expect that upon increasing $\bce$, 
$\tcar$  decreases monotonically from  $\tcarmax$ to $\tcarmin$.   
To simplify even further, we use the fact that $\taure$ is often small 
compared to all others time involved in the problem, i.e. $\taure \ll \taurp, \tauce, \taucp$.   
Then
\be
\tcarmax  &=&       \taucp \left[ 1 + \frac{\taure}{\taurp} + O (\taure^2)   \right],  \label{eq:tcarmaxtoy} \\
\tcarmin  &=&      \frac{\taurp \taucp}{\taurp + \taucp} + O (\taure).                 \label{eq:tcarmintoy}
\ee
Compared to Eq.~\ref{eq:tcartoy}, 
both expressions are numerically accurate
and no further term in the expansion is necessary. 
As regards the ratio $r$, the above approximations yields
\be
r= \frac{\tcarmin}{\tcarmax} &=&       \frac{1}{(1+\taure/\taurp) (1+\taucp/\taurp)},                        \nonumber   \\
                              &\simeq&  \frac{1}{ 1+\taucp/\taurp}   \qquad \mbox{for\ \ } \taure \ll \taurp.   \nonumber
\ee
This shows that at low $\bc$ values, $\tcar$ depend largely on $\bce$. 
For instance, taking $\bc=10^7$~\footnote{We remind that conductances and other physical quantities are expressed in SI units, unless otherwise indicated.} yields $r\simeq1/100$, i.e. one obtains a 100-fold reduction in $\tcar$
when $\bce$ increases from $0$ to $\infty$. 

For conductances $\bc$ smaller than $10^8$, $\tcar$ is actually rather well approximated by the 
simple expression
\be
\frac{\tcar - \tcarmin}{\tcarmax - \tcarmin} &=& \frac{1}{1+\bce/\bc}.    
\label{eq:tcarapptoy}
\ee
Deviations amount to only a few percents for $\bc < 10^8$, 
and occur only at very large $\bc$, 
where $\bc$ should be replaced with a function $g(\bc)$ which increases slightly less than $\bc$ (see below). 
The resulting cooling time is shown in Fig.~\ref{fig:res-2Tfilm}, which demonstrates a strong effect of the 
electron-phonon interfacial conductance $\bce$, especially for low values of $\bc$.
A useful approximation for the following may be obtained, if we consider that  $\tcarmin \simeq  \frac{\taurp \taucp}{\taurp + \taucp} \simeq \taurp$
and  $\tcarmax \simeq \taucp$, which is a reasonable approximation given that $\taucp \gg \taurp$ with the default parameters.
This yields
\begin{equation}
\tcar \simeq \taurp + \frac{\taucp}{1+\bce/\bc}, 
\label{eq:tcarapp2toy}
\end{equation}



%
\subsubsection{Full two-temperature model}
The FD model neglects diffusion of the energy carriers in the metal, which may induce additional resistance to the heat flow.
Here, we go one step further as compared with the FD model, and account explicitely for the electron and phonon diffusion. 
As detailed in Appendix~A, 
the long-time behavior of the temperature decay depends on whether 
the substrate temperature is constant or not.
In the latter case, 
the decay is power-law, without any characteristic time scale. 
In the former case,
the decay may be characterized by a single decay time $\tcar$. 
This is the reason why we use this cold substrate approximation
and  determine analytically how the characteristic decay time $\tcar$ depends on $\bc$ and $\bce$. 
We find that if $ \bce \gg \bc$, the temperature decay is much faster than that expected for zero $\bce$. \newline
Using $h$ and $\taure$ as the length and time units, the relevant equations, boundary and initial conditions read as:
\be
\begin{array}{rlrlrrl}
     \partial_t \Te &  \multicolumn{5}{l}{ =  \frac{1}{\Me} \; \partial_{xx}^2  \Te - (\Te-\Tp), }                 \\ 
R \, \partial_t \Tp &  \multicolumn{5}{l}{ =  \frac{1}{\Mp} \; \partial_{xx}^2  \Tp + (\Te-\Tp), }                 \nonumber  \\
    \partial_x \Te  & = 0 ,             &      \partial_x \Tp & = 0,       & \mbox{\ \ \ \ \ for\ \ }&  x &= 0,   \nonumber  \\
  - \partial_x \Te  & = \bie \Te,       &    - \partial_x \Tp & = \bip \Tp,  & \mbox{for\ \ }&  x &= 1,             \nonumber  \\
               \Te  & = T^i,  &                 \Tp & = 0,        & \mbox{for\ \ }&  t &= 0.      \label{eq:full_two_temperature_model}
\end{array}
\ee
%
There are thus six dimensionless parameters: $\bie$, $\bip$, $\Me$, $\Mp$, $R$   and $\delta/h$.
The general solution of the previous set of equations is not exponential. However, in order to characterize
the long time decay, we consider Laplace transforms with respect to time and take its expansion close to zero frequency. As further explained in the appendices A and B, this gives an estimate of the longest relaxation time characterizing the long time cooling kinetics. Even if this relies on an approximation, we believe that it can reflect faithfully the influence of $\bce$.

\begin{figure}
\includegraphics[width=8.5cm]{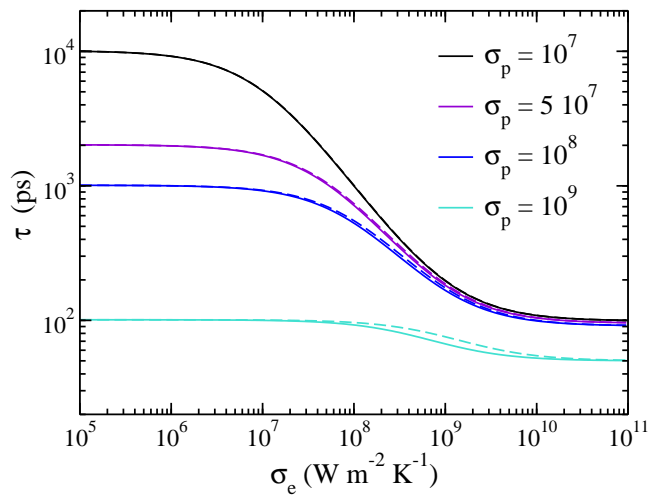}
\caption{Characteristic decay time $\tcar$  as a function of the electron-phonon conductance $\bce$, 
for several phonon-phonon conductance $\bc$, 
as computed with the FD model and with the full two-temperature (dashed and solid lines respectively).}
\label{fig:res-2Tfilm}
\end{figure}

For the two-temperature model considered here, 
the characteristic decay time $\tcar$  can be obtained analytically 
using a symbolic computation software, 
but the expression is much too large to be reported. 
As shown in Fig.~\ref{fig:res-2Tfilm} for the default parameters and a wide range of $\bc$ and $\bce$, 
large $\bce$  can result in a significant decrease of the decay time. Alternately, this trend may be captured analytically 
as detailed in the Appendix B. 
Finally, as visible in Fig.~\ref{fig:res-2Tfilm}, 
the FD model provides a very reasonable approximation to the 
full two-temperature model, 
showing that even when ``neglecting'' diffusion, one can recover the qualitative behavior.  This suggests that 
the effective conductance characterizing the different energy channels present at a metal-non metal interface, does not depend primarily on the diffusion of the energy carriers in the metal.

\subsection{Effective thermal boundary conductance}
From the solution of the two models analyzed above, we can predict the effect of the different electron-phonon processes on 
the effective thermal boundary conductance $\sigma_{\rm eff}$. This latter quantity is the apparent interfacial conductance 
extracted from a TTR signal using a one-temperature model. Because the one-temperature model ignores the electronic degrees of freedom, 
the effective conductance depends on both the electron-phonon coupling constant and the conductance $\bce$. Here, we discuss this dependence.
The effective conductance is computed using
\begin{equation}
\sigma_{\rm eff}=q/(\Tp-T_0),
\label{effective_conductance}
\end{equation}
where $q=-h \frac{d \Tp}{dt}$ is the interfacial heat flux density. Note that here we are assuming the phonon temperature to be uniform, in the spirit of the FD model as we concluded that phonon diffusion within the metal film does not change significantly the metal cooling kinetics.


Because in the FD model discussed previously, the long time decay of the metal lattice temperature is exponential, with a time constant $\tcar$ the effective conductance Eq~(\ref{effective_conductance}) is given by: $\sigma_{\rm eff}=\cp h/\tcar$. Using Eq.~(\ref{eq:tcarapp2toy}) for the characteristic time, we predict the following expression of the conductance:
\begin{equation}
1/\sigma_{\rm eff}=1/Gh +1/(\bce + \bc).
\label{effective_conductance_2}
\end{equation}
This equation has a simple interpretation: the effect of the electron-phonon processes is
equivalent to add in series the bulk resistance $1/Gh$ with the total interfacial resistance $1/(\bce+\bc)$. This latter resistance  comprises two parallel interfacial resistances quantifying electron-phonon and phonon-phonon processes at the interface. At this point, it is important to realize that the expression of the bulk resistance $1/Gh$ differs from the expression $1/\sqrt{G \kp}$ derived by Majumdar and Reddy~\cite{majumdar2004}. The difference stems from the different boundary conditions considered for the electrons: Ref.~\cite{majumdar2004} assumed an insulating boundary condition for the electrons, as they were not allowed to directly exchange energy with the substrate. Here, because we consider the possibility of a direct exchange at the interface, the boundary condition is not adiabatic. This difference 
of boundary condition changes the expression of the internal resistance governed by electron-phonon processes in the metal thin film. 

Finally, we discuss the traditional separation between equilibrium and non-equilibrium contributions to the effective conductance~\cite{majumdar2004,wilson2013}. The expression of the conductance Eq.~\ref{effective_conductance_2} has been derived under non-equilibrium conditions. If we had assumed equilibrium, we would have obtained the intuitive expression of the interfacial conductance 
$\bce+\bc$ which expresses the fact that the two energy channels$\;$-electrons and phonons-$\;$act in parallel to transmit heat to the substrate.  
Hence, we find that the non-equilibrium contribution consistent with an isothermal boundary condition is 
\begin{equation}
\sigma_{\rm neq} = Gh,
\label{non_equilibrium_conductance}
\end{equation}
and is found to be larger for thick metal films $h>100$ nm, and metals having a large electron-phonon coupling $G >10^{17}$ W m$^{-3}$ K$^{-1}$. 
For these systems, the non-equilibrium conductance is $\sigma_{\rm neq}>10^{10}$ W m$^{-2}$ K$^{-1}$ and the corresponding resistance is practically 
negligible, as compared with the phonon-phonon and electron-phonon interfacial resistances. 
On the other hand, for thin Au films with $h<10$ nm the non-equilibrium conductance is $\sigma_{\rm neq} < 250$ MW m$^{-2}$ K$^{-1}$ and may become 
comparable with the interfacial conductances $\bc$ and $\bce$. We conclude then that in metal-non metal superlattices with a small period, 
non-equilibrium effects can not be neglected, and should contribute to enhance the thermal resistance.

\begin{figure}[htb]
\includegraphics[width=8cm]{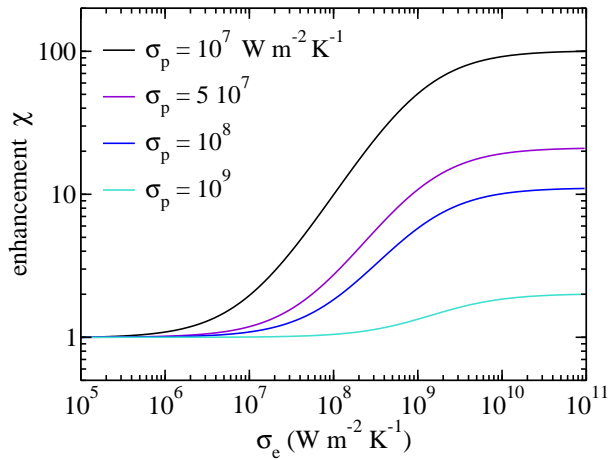}
\caption{
Relative enhancement $\chi$ of the apparent thermal boundary conductance 
due to the electron-phonon conductance $\bce$, as predicted by the fast diffusion approximation  
Eq.~\ref{effective_conductance_2} with default parameters. 
}
\label{fig:enhancement_toy}
\end{figure}

We now discuss the effect of a finite $\bce$ on the value of the effective conductance Eq.~(\ref{effective_conductance_2}).
All the expressions of the characteristic time derived so far clearly indicate
that the electron-phonon conductance $\bce$ has a strong effect on the long time decay of the temperature of the metal, 
and may significantly accelerate the cooling kinetics. 
To further illustrate this point, 
we have considered the relative enhancement of the metal cooling induced by the electron-phonon conductance. 
This enhancement is simply defined by the ratio $\chi=\sigma_{\rm eff}(\bce)/\sigma_{\rm eff}(\bce=0)$ and as such may be considered 
as a simple estimate of the increase of the apparent thermal boundary conductance induced by $\bce$. 
Figure~\ref{fig:enhancement_toy} shows that the relative enhancement 
may reach one or two orders of magnitude, 
depending on the value of the phonon-phonon conductance $\bc$. 
The enhancement is most significant when the electron-phonon conductance $\bce$ becomes greater than $\bc$, 
a situation that might be common, as discussed in Sec.~\ref{sec:motivations}. 
Again, we conclude that the largest enhancements are observed when the metal/substrate interface transmits poorly the phonons, which corresponds to the case of acoustically mismatched solids. 


\section{Two realistic cases: Gold/silicon and gold/silica interfaces}
\label{sec:realistic}

\subsection{Methods and parameters}
We have discussed above the effect of $\bce$ using a set of typical values for the different transport and thermodynamic 
quantities characterizing the metal and the dielectric. 
Here, we consider two realistic cases, 
which allows to take into account the temperature dependence of the transport coefficients. 
We focus on the gold/silicon and gold/silica systems, 
for which the values of the electron-phonon interfacial conductance are known~\cite{hopkins2009}

\begin{table}[h]
\begin{tabular}{lccccccc}
\hline \hline  
  &  $k$  &  C$_p$ & $T_{\rm Debye}$ &  $n$ & $\bce $ & $\bc$                \\
\hline 
Gold phonons     & 18.0  &  2.35 $\times$ 10$^6$  &  165 &  97970 &  -     & -       \\
Gold electrons   & 300.0  &  1.97 $\times$ 10$^4$  &    - &  97970 &  -     & -       \\
Silicon Si       & 153.6  &  1.68 $\times$ 10$^6$  &  625 &  82891 &  134   &  72.6   \\
Silica SiO$_2$   & 1.370  &  1.01 $\times$ 10$^6$  &  403 &  44075 &  152   &  141.5  \\
\hline  \hline
\end{tabular}
\caption{Thermophysical parameters at 300 K: 
thermal conductivity $k$ (W\,m$^{-1}$\,K$^{-1}$),
heat capacity C$_p$ (J\,m$^{-3}$\,K$^{-1}$), 
Debye temperature $T_{\rm Debye}$ (K), 
molar density $n$ (mol\,m$^{-3}$),   
electron-phonon $\bce$ and phonon conductance $\bc$ (\bcunitM).}
\label{Table_parametres_300K}
\end{table}

The electron-phonon coupling factor was set to the value $G=2.5 \times 10^{10}$ MW\,m$^{-3}$\,K$^{-1}$
predicted by DFT calculations when the electronic temperature is below $5000$~K~\cite{lin2008}. 
The values of the other thermophysical parameters at $300$ K are given in Tab.~\ref{Table_parametres_300K}.   
The heat capacity of the gold electrons is calculated with 
$C_\el = \gamma T_\el$ where $\gamma=65.64$  J\,m$^{-3}$\,K$^{-2}$ is Sommerfeld's constant of gold,  
while the vibrational heat capacity is obtained from standard expression~\cite{AshcroftMermin}.
As regards the thermal conductivity of the substrate, 
we used experimental values for silicon and silica extracted 
from Refs.~\cite{Glassbrenner1964} and \cite{Jund1999} respectively, as briefly described in the Appendix~C.
Finally, we take the electronic thermal conductivity and the phonon thermal conductivity to be constant in gold metal. 

The phonon-phonon thermal boundary conductances are assumed 
to be described by the diffuse mismatch model,  
as discussed in Sec.~\ref{sec:motivations}. 
Because gold metal has a low Debye temperature $\Theta_D =165$ K, 
we can assume that above $200$ K, the phonon\itf phonon conductance is constant, and takes the value
\begin{equation}
\label{DMM_high_T}
\bc = \frac{n k_B}{4} \left( \sum_p v_{1,p} \right) \alpha, 
\end{equation}
where $n$ is the number density, and the phonon transmission coefficient $\alpha$ has been calculated in eq.~\ref{transmission_coefficient}. 
Numerically, this yields  
$\bc=72.6$  \bcunitM  for the Au/Si interface, and 
$\bc=141.5$ \bcunitM  for the Au/Si$O_2$ interface. 

The temperature dependence of the electron-phonon conductance for silica and silicon
was extracted from Ref.~\cite{hopkins2009}, 
using  the following expressions
\begin{eqnarray}
\label{eq_fit_hep}
\bce^{\rm Au/Si}    &=&   25.18 + 0.363\  T,  \label{bcpexpsilicon}  \\
\bce^{\rm Au/SiO_2} &=&   96.12 + 0.189\  T,  \label{bcpexpsilica}
\end{eqnarray}
where temperature is given in~K, and conductances in \bcunitM. 
Equations~\ref{bcpexpsilicon} and ~\ref{bcpexpsilica} provide a reasonable description of the experimental data, 
as shown in Fig.~\ref{fig_electron_phonon_conductance}. 

The equations of the two-temperature model are solved using a finite-difference scheme, 
with perfectly matched layers at the boundaries so as to simulate a quasi-infinite substrate, whose thickness is fixed at 50~$\mu$m.  
Gradients are estimated using finite differences with three successive lattice points, while the condition $h (T(x^{+},t)-T(x^{-},t)) = k  \partial_x T$ is enforced at the interface using two points to evaluate the local gradient. 
The typical lattice step is $5$ nm; the timestep is chosen so that the maximal temperature increment between two successive time steps is $\Delta T=\max_x \vert T(x,t_{n+1})-T(x,t_{n}) \vert =0.01$ K if $t<1$ ps; $\Delta T=0.5$ K otherwise unless the temperature difference becomes $\vert T-T_0 \vert < 2$ K, when a finer criterion $\Delta T=0.05$ K is retained. 
Here and in all the following $T_0$ is the initial temperature of the system, prior to any heating. 
The initial value of the electronic temperature is $T_e=T_0$ + $2000$~K,  
the metal penetration depth is $\delta=20$~nm and 
the film thickness is fixed to $h=100$~nm. 


\subsection{Results}
We present now the results obtained by numerically solving the two-temperature model, 
using the physical parameters detailed above. 
As in Sec.~\ref{sec:anal}, 
we focus on the electronic temperature of the metal at the air/metal surface, 
since this is a simple measure of the metal reflectivity, 
which is probed in the time-resolved thermoreflectance experiments. 

\begin{figure}[h]
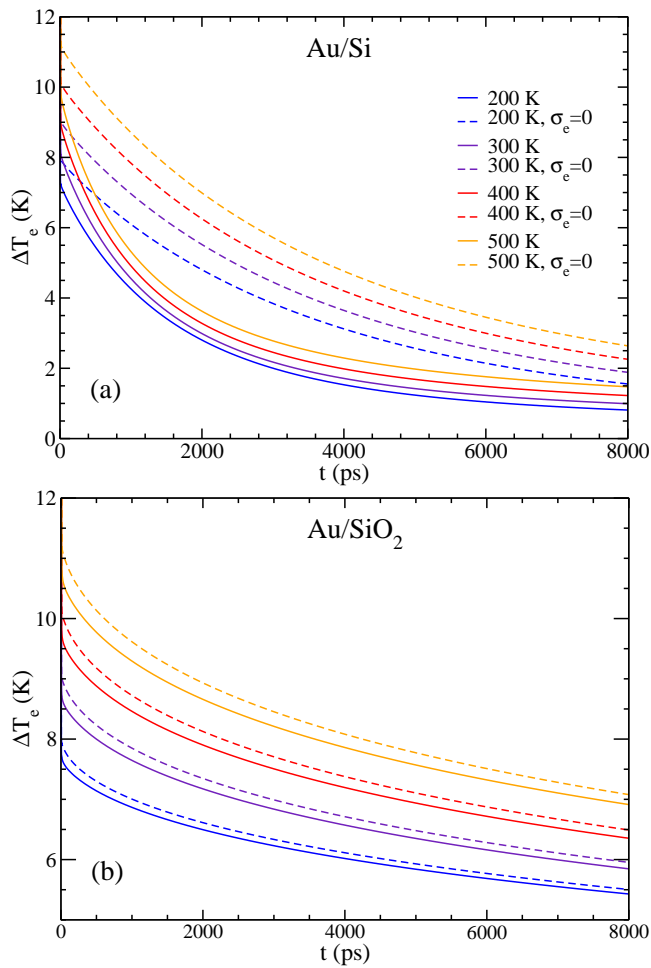

\includegraphics[scale=0.31]{Fig8a.eps} \\
\includegraphics[scale=0.31]{Fig8b.eps}
\caption{(Color online) Effect of the electron-phonon interfacial conductance  $\bce$
on the cooling of gold metal's electrons, 
as found with the two-temperature model for gold film on silicon substrate (a) 
and on silica substrate (b).  
Solid lines: simulations with $\bce$ given by Eqs.~\ref{bcpexpsilicon} and \ref{bcpexpsilica} ; 
Dashed lines: same simulations with $\bce=0$.}
\label{fig_cooling_T}
\end{figure}

\begin{figure}[h]
\includegraphics[scale=0.31]{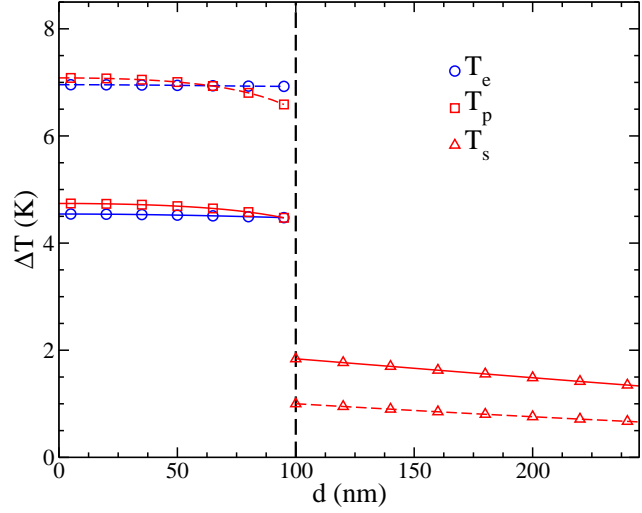} \\
\caption{(Color online) Electron and phonon temperature profiles across the gold/silicon interface, $1000$ ps after initial heating. Solid lines correspond to simulations with $\bce$ given by Eq.~\ref{bcpexpsilicon}, while dashed lines are simulations 
without interfacial electron-phonon coupling $\bce=0$. The temperature profiles of the substrate are also shown. 
}
\label{fig_profiles_silicon}
\end{figure}

Figure~\ref{fig_cooling_T}.a displays the temporal evolution 
of the surface electron temperature for the gold/silicon  system. 
For each temperature, curves with the  electron-phonon conductance switched off ($\bce=0$) are also shown. 
The difference between the two cases is striking. 
Clearly, discarding the electron-phonon channel results in  an electron cooling which is much 
slower. After $~1000$ ps, the electronic temperature is almost twice larger in the case of zero $\bce$.
Figure~\ref{fig_cooling_T}.a thus illustrates the key point of this work:
in the time interval relevant to thermoreflectance measurements $100-1000$ ps, 
the electron-phonon interfacial coupling has a strong effect on the temperature decay. 
Therefore it should be taken into account, along with the phonon-phonon conductance.  
It should be acknowledged at this point, that for the nanosecond time regime we are considering here, electrons and phonons 
in the metal are practically at equilibrium as illustrated in Fig.~\ref{fig_profiles_silicon}. A slight difference between electron 
and phonon temperature is however visible when $\bce=0$, as a result of the adiabatic boundary condition obeyed by the electrons in this case.

\begin{figure}[h]
\includegraphics[scale=0.31]{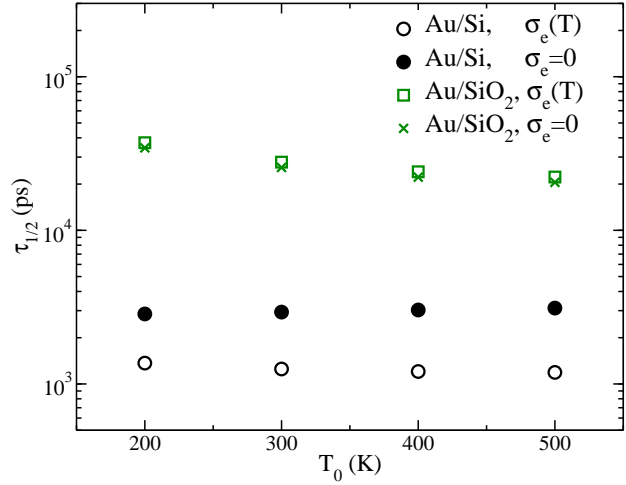}
\caption{(Color online) Cooling time $\tauh$  (see text) for the gold/silicon and gold/silica systems, 
as a function of the temperature $T_0$ prior to heating. 
Open symbols are simulations where $\bce$ given by eqs.~\ref{bcpexpsilicon} and \ref{bcpexpsilica}, 
while crosses and stars correspond to the case $\bce=0$.  
}
\label{fig_tau_T_infinite}
\end{figure}

To quantify the effect of $\bce$ on the temperature decay, 
we introduce a time $\tauh$ defined as the time after which the electronic temperature has cooled down to half its initial value
\footnote{At early times, electrons may transfer their excess energies to the substrate, which results in an early time electronic 
temperature which might depend on the value of the electron-phonon boundary conductance.}. 
$\tauh$ is a simple estimate characterizing the cooling kinetics, that may be used in experiments. 
%
Figure~\ref{fig_tau_T_infinite} displays $\tauh$ as a function of the initial temperature $T_0$. 
The effect of the interfacial electron-phonon coupling is clearly seen: 
in presence of this channel, 
the electron cooling is typically a factor two to three faster than in its absence. 
In terms of apparent thermal boundary conductance, 
this corresponds to a change by an amount ranging from $200$ to $300 \% $. 
Again, we see that interfacial electron-phonon coupling should contribute significantly 
to the cooling process and to the apparent value of the thermal boundary conductance.
Note also that Fig.~\ref{fig_tau_T_infinite}  shows cooling times that are practically independent of the initial temperature $T_0$, 
in both cases where $\bce$ has been set to zero or not.
This may certainly be explained by the use of a constant phonon-phonon conductance in the model.
The use of a temperature dependent electron-phonon conductance $\bce$  does not change the dependence on $T_0$.

We have also characterized the effect of $\bce$ on the cooling kinetics of the gold/silica system. 
As  illustrated in Fig.~\ref{fig_cooling_T}.b,  
the temperature decay is weakly dependent on the interfacial electron-phonon conductance.  
This is confirmed by Fig.~\ref{fig_tau_T_infinite} where  $\tauh$ is seen to be practically 
insensitive to the presence of the additional interfacial energy channel. 
This relative difference between the cases of gold/silicon and gold/silica may be explained by two factors. 
The first is the difference in relative magnitude of the two conductances: 
while $\bce$ is twice larger than $\bc$  for gold/silicon,  
they are nearly equal in the case of gold/silica.  
The other factor is the difference of thermal conductivities of the substrates, 
with  silicon  having a conductivity two orders of magnitude higher than silica.   
As a result, any energy flux coming from the metal's electrons is conveyed away from the interface. 
In contrast, heat in the silica remains confined in the vicinity of the interface, 
slowing down any energy transfer coming from the metal's electrons, 
and resulting in the relative insensivity to $\bce$ in the cooling kinetics of the gold/silica system.

\begin{figure}[h]
\includegraphics[scale=0.28]{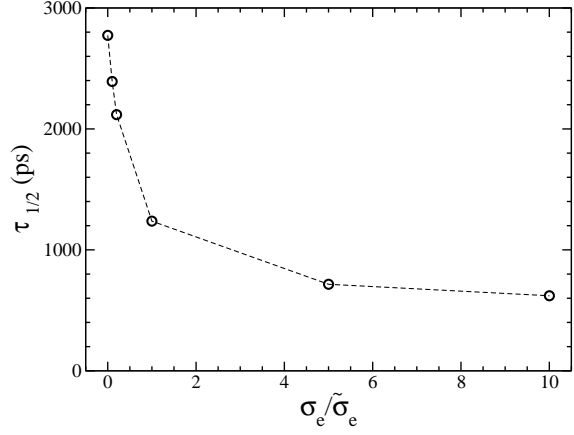}
\caption{Effect of the electron-phonon conductance $\bce$ 
on the  cooling time $\tauh$ (see text) 
for the gold/silicon system.  
Here $\tilde{\bce}$ denotes the value extracted from experiments~\cite{hopkins2009} and given by Eq.~\ref{eq_fit_hep}. 
The temperature is  $T_0=300$ K.
The line is a guide to the eye.  
}
\label{fig_tau_hep}
\end{figure}

As a final illustration of the relevance of the electron-phonon boundary conductance, 
we  consider for the  case of a gold/silicon interface 
a conductance $\bce$ varying in a wide range. 
The time $\tauh$ is shown in Fig.~\ref{fig_tau_hep}, 
where the abscissa axis has been normalized by the 'experimental' value given by Eq.~\ref{eq_fit_hep}.
The decrease of the cooling time $\tauh$ is very sharp for low $\bce$, 
suggesting that even a moderate electron/substrate coupling may lead to an enhanced thermal conductance. 
For larger values of $\bce$, the decrease is somewhat less pronounced, 
but still leading to a threefold faster cooling rate for the highest value studied. 
Again, this roughly corresponds to a factor three in the apparent thermal boundary resistance.  
Note that the corresponding value of $\bce \simeq 1300$ \bcunitM 
may be realistic to describe metals other than gold, 
as we discussed in Sec.~\ref{sec:motivations}.

\section{Conclusion}
\label{sec:con}
In summary, we used a combination of analytical work and numerical simulations of the two-temperature model 
to probe the influence of a direct interfacial electron-phonon energy transfer 
in the thermal transport between a metal, or a semi-metal and a dielectric substrate.
We have shown that Sergeev's model predicts  for most  metal-non metal systems considered here 
an electron-phonon conductance~$\bce$  
which is larger or comparable with the phonon-phonon conductance~$\bc$. 
Perhaps surprisingly, even semi-metals such as Bismuth, with poor electronic transport properties  compared to those of noble metals, 
may involve a significant~$\bce$, 
but which remains smaller than~$\bc$.   
This suggests that electrons of materials which are poor electronic conductors may participate in interfacial heat transport. 
 
To assess the effect of a finite electron-phonon interfacial conductance, 
we investigated the cooling 
of a metal thin film instantaneously heated by a strong laser pulse. 
This situation is of direct relevance to time-resolved thermoreflectance technique, 
which is used to measure the thermal boundary resistance between the metal and the substrate.
The effective interfacial conductance consists in a non-equilibrium contribution $Gh$ and 
an equilibrium conductance $\bc+\bce$. The non-equilibrium conductance is found to be important for 
metals having a relatively low value of $G$, and for thin films a situation relevant to metal-dielectric superlattices 
or core-shell nanoparticles. The effect of $\bce$ is found to enhance the apparent thermal boundary conductance
as compared with $\bc$, up to two orders of magnitude for acoustic mismatched interfaces, 
characterized by low values of $\bc$ on the order of $10$ \bcunitM. This may explain the large discrepancies reported 
between the apparent conductances measured for solids having a large Debye temperature ratio and the classical DMM model~\cite{stevens2005}.
%
%
Numerical simulations of the two-temperature model on two realistic cases 
confirm  our analytical findings: 
while $\bce$ has only a mild effect on the cooling of gold/silica system, 
it significantly speeds up the cooling of the gold/silicon system.  
Those distinct behaviors originate in the difference in the phonon-phonon conductance $\bc$ and  the thermal conductivities of the substrate.

On the theoretical side, one final remark is in order. We emphasized the strong limitations of Sergeev's model to predict the electron-phonon 
conductance $\bce$. In particular, Sergeev's model disregards the nature of the substrate, in disagreement with the experimental results~\cite{hopkins2009}, and seems to overestimate the interfacial electron-phonon conductance, 
in the few cases when comparison with experiments is possible.  
Hence, there is a clear need to develop theoretical models for the electron-phonon conductance 
that would account for the nature of the non-metal substrate. 
Given the small number of $\bce$ values that have been obtained experimentally so far, 
measurements for a variety of metal/substrate systems would be invaluable in this endeavour~\cite{giri2014}.


\subsection*{Appendix A: Characterizing the long-time temperature decay.}
\label{sec:expvspl}


This appendix focuses on the long-time behavior of the temperature decay. 
We find that when the substrate heating is negligible, 
which we call the ``cold substrate'' approximation,  
the long-time decay may be characterized by a single decay time~$\tcar$. 
On the other hand, if the substrate heating is significant the long-time decay is power-law, with no characteristic time. 
This distinction is essential when one wants to extract the thermal boundary conductance from the experimental curves. 
For simplicity, we use the simple one-temperature model to make this point,  
but it applies to the two-temperature model as well.  

\paragraph*{Cold substrate. }
Here we assume that the temperature increase of the substrate is negligible, 
that is $\Ts = 0$ at all time. 
Taking $h$ as the unit length 
and the phonon conductance time  $\taucp=\cp  h/\bc$ as the unit time, 
the adimensionalized equations for the one-temperature model are
\be
\begin{array}{rllrrl}
\bip \,      \partial_t \Tp &  \multicolumn{4}{l}{ =   \; \partial_{xx}^2  \Tp , }                 \nonumber  \\
             \partial_x \Tp & = 0 \mbox{\ \ \ \ \ \ \ \ }           & \mbox{for\ }&  x &= 0,             \nonumber  \\
 \bip \Tp  + \partial_x \Tp & = 0              & \mbox{for\ }&  x &= 1,             \nonumber  \\
                        \Tp & = T^i(x)         & \mbox{for\ }&  t &= 0.          
\end{array}
\ee
There are only two parameters: 
the Biot number $\bip=h \bc/\kp$  and the $\delta/h$ ratio. 
We now introduce $\Ttip(x,s)$, the Laplace transform of $\Tp$ with respect to time, which satisfies the equation
\be
 -\partial_{xx}^2 \Ttip + \bip \Ttip  &=&   \bip T^i(x),  \nonumber
\ee
whose solution is 
\be
\Ttip(x,s)  &=&  C_1 e^{\gamma x} +   C_2 e^{-\gamma x} + \Tti_{part}, \qquad \gamma = \sqrt{\bip s},        \nonumber \\
\Tti_{part} &=& \frac{\bip}{\gamma} \; \int_0^x \sinh \left[ \gamma (u-x) \right] T^i(u) du.                  \nonumber
\ee
The constants $C_1$ and $C_2$ are obtained from the boundary conditions at $x=0$ and $x=1$. 
From now on, we focus on the temperature at the air\itf metal interface $\Ttip(s)= \Ttip(0,s)$, 
and consider the long-time behavior by taking a small-$s$ expansion.
The result is $\Ttip(s) \simeq a_0 + a_1 s + O(s^2)$.
Note that for a time dependent function $f(t)$, whose Laplace transform has the expansion $f(s)= a_0 + a_1 s + O(s^2)$,
we can always characterize the long time decay by an effective time $\teff = -a_1/a_0$. 
If  $f(t)\sim e^{-t/\tau}$ is a single exponential, one recovers $\teff=\tau$. 
If on the other  hand, $f(t)$ is a sum of exponential terms:
\be
\Ttip(t) = \sum_n c_n\, e^{-t/\tau_n},
\ee
then the procedure yields
\be
\teff = \frac{\langle \tau _n^2 \rangle }{ \langle \tau_n \rangle},  \quad  \langle  x \rangle \equiv \frac{\sum c_n x_n}{\sum c_n },  
\ee
that is an average that takes into account the amplitude of each mode, 
and gives more weight to the slow modes.

\begin{figure}[htb]
\includegraphics[width=7cm]{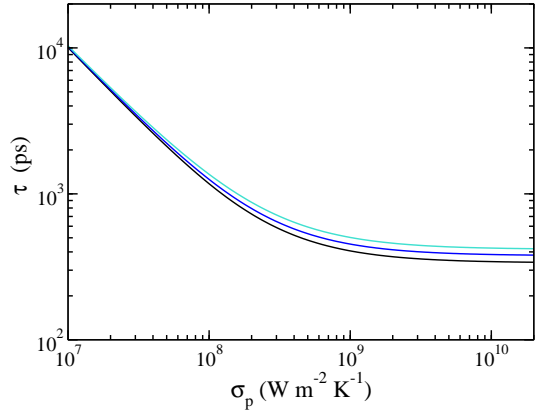}
\caption{Characteristic decay time $\tcar$ in the one-temperature model 
as a function of the conductance $\bc$, 
with default parameters, and penetration depth $\delta=0$, $20$~nm and $\infty$, from bottom to top.}
\label{fig:an1Tfilm}
\end{figure}

Following this procedure for the metal phonon temperature, we obtain 
the characteristic time $\tcar$ for arbitrary penetration depth $\delta$ 
\begin{widetext} 
\be
\tcar  &=& \frac{3 (\bip \delta -1) \left(\bip \left(2 \delta ^2-1\right)-2\right)-2
   e^{\frac{1}{\delta }} \left(\bip \left(3 \bip \delta ^3-3 (\bip+1) \delta
   ^2+\bip+3\right)+3\right)}{6 \left(e^{\frac{1}{\delta }} (\bip (\delta -1)-1)-\bip
   \delta +1\right)},
\label{eq:1Ttaudeltaadim}  \\
\tcar  &=& -\frac{-2 e^{h/\delta } \left(\kappa ^2 \left(h^3-3 h \delta ^2+3 \delta ^3\right)+3 \kappa 
   (h-\delta ) (h+\delta )+3 h\right)-3 (\delta  \kappa -1) \left(h^2 \kappa +2 h-2 \delta ^2 \kappa
   \right)}{6 \alp  \kappa  \left(e^{h/\delta } (h \kappa -\delta  \kappa +1)+\delta  \kappa
   -1\right)}.  
\label{eq:1Ttaudelta}
\ee
\end{widetext}
Equation~\ref{eq:1Ttaudeltaadim} applies if length and time units are $h$ and $\taucp$ respectively.  
Equation~\ref{eq:1Ttaudelta} is given in terms of dimensional parameters, 
and for simplicity of notations, we have introduced the inverse length $\kappa=\bc/\kp$. 
This formula simplifies in the two limiting cases
\be
\delta \rightarrow 0,       \quad\quad   \frac{\tcar}{\taucp} &=&  1  +  \frac{\bip^2          }{3(\bip +1)}    ,   \\
\delta \rightarrow \infty,  \quad\quad   \frac{\tcar}{\taucp} &=&  1  +  \frac{\bip (5 \bip + 8)}{12(\bip +2)}   .   
\ee
The Biot number $\bip$ indicates the relative importance between the 
heat flux within the bulk of the film 
and the flux at the interface. 
If $\bip \ll 1$, the temperature is quasi-uniform within the film 
and the flux is limited by the interface resistance, 
then $\tcar=\taucp$, whatever the initial profile. 
On the other hand, if $\bip \gg 1$, 
diffusion within the film  is the limiting factor 
and $\tcar \sim  h^2/\alp = \taucp$ 
is the typical time of phonon diffusion over the film thickness~$h$, 
with a  prefactor  that depends on the initial profile. 
Both limiting cases are visible in Fig.~\ref{fig:an1Tfilm}, 
which shows the characteristic time $\tcar$ as a function of $\bc$. 
Interestingly, with the default parameters, 
a typical value $\bc=10^8$ yields $\bip=1$, 
meaning that none of the limits applies. 
And indeed, the difference is significant between $\taucp=1000$ ps 
and $\tcar=$1167, 1256 and 1361~ps for $\delta=0,20,\infty$ respectively. 

\paragraph*{Heated substrate. }
Retracing the steps of the analysis above, 
one finds for the  small-$s$ expansion of the surface temperature
\be
\Ttip(s) &=& \frac{a_{-1/2}}{\sqrt{s}} + a_0 + a_{1/2}\sqrt{s}+  a_1 s + a_{3/2} s^{\frac{3}{2}}  + O(s^2),  
\ee
When taking the inverse Laplace transform,  
the terms with half-integer exponents yield power-laws, 
which are dominant over other terms~\footnote{The terms $ a_0 +  a_1 s$ yields an exponential $e^{-t/\tau_a}$, with $\tau_a=-a_1/a_0$. For $t>\tau_a$, it becomes quickly negligible with respect to the long-time tails.}.  
We thus expect for $\Tp(t)$ the approximation
\be
\Tp(t) &=& \frac{a_{-1/2}}{\sqrt{\pi}}\, t^{-\frac{1}{2}} - \frac{a_{1/2}}{2\sqrt{\pi}}\, t^{-\frac{3}{2}} 
         +  \frac{3\,a_{3/2}}{4\sqrt{\pi}}\, t^{-\frac{5}{2}} +  O(t^{-\frac{7}{2}})
\label{eq:powerlaw}
\ee
The key point is that the decay is power-law, since at the longest time, $\Tp(t) \simeq a_{-1/2}/\sqrt{ \pi t}$.
This is a general feature of diffusion on a half-line~\cite{Redner-FirstPassage2001}.  
Furthermore, the coefficient  $a_{-1/2} = h \cp/\sqrt{\cs \ks}$ 
depends  on the total energy injected  and on the substrate properties, 
but not on the interfacial conductance $\bc$~\footnote{The coefficients  $a_{1/2}$,   $a_{3/2}$, and subsequent do depend on $\bc$.}. 
In contrast with the cold substrate approximation treated above, 
it is not possible to define a single time scale that could characterize the temperature decay. 
Thus, those calculations with the one-temperature model show that 
two cases should be distinguished. 
If substrate heating is negligible, the decay may be characterized by an effective time $\tcar$, and presumably the decay time $\tcar$ can be easily extracted, 
from which, if all other parameters are known, $\bc$ can be deduced. 
If on the other hand, the cold substrate approximation does not apply, 
the temperature decay is power-law, and at the longest time, completely independent of $\bc$. 
It might be possible, using fits of the experimental curve, to estimate the coefficients  $a_{1/2}$,   $a_{3/2}$, and subsequent, 
and deduce $\bc$, but the process seems much more delicate. 
This conclusion applies {\it a fortiori} to the two-temperature model.

\subsection*{Appendix B: Two-temperature model in the cold substrate approximation.}
Here, we apply to the full two-temperature model the analysis detailed in Appendix A.
To do so, let's introduce the dimensionless parameters
\be
a^2 &=&  \Me + \Mp ,                                        \\
b^2 &=& \frac{ (\Me^2 + R \Mp^2 )^2  }{ 4(\Me + \Mp)^3},    \\ 
c^2 &=& \frac{\Me  \Mp \, (1+R)}{\Me + \Mp}.                                                                                       
\ee
The full expression for the characteristic time $\tcar$ is very unwieldy, 
in part because $\tcar$ involves power of $e^a$, 
an exponential term can not be easily approximated since 
$a$ is neither small nor large:  
with the default parameters, $a=3.31$ is of order unity. 

To get a qualitative understanding of the $\tcar$ dependence on $\bce$, 
we first consider the two limiting values for  
$\bce \rightarrow 0$ and 
$\bce \rightarrow \infty$, 
which are denoted as $\tcarmax$ and $\tcarmin$. 
Even after taking the limits, the expressions for $\tcarmax$ and $\tcarmin$ are still untractable. 
We therefore resort to two bold approximations. 
First,  the limit $\delta/h \rightarrow 0$ is taken, even if the value $0.2$ is not so small.  
Second, a large-$a$ expansion is made, even though $e^a \simeq 27$ is not so large. 
The resulting expressions are quite simple 
\be
\frac{\tcarmax}{\taure}  &=&   1 +  \frac{c^2}{\Me}   \left[ \frac{a^2}{\bip} + a -1 + O\left( \frac{1}{a}\right) \right], 
\label{eq:tcarmax}  \\
\frac{\tcarmin}{\taure}  &=&  \frac{c^2}{3} +  \frac{6 b + c^2}{3 a} -  \frac{3 b + 2 c^2}{3 a^2} + O\left( \frac{1}{a^4}\right),  
\label{eq:tcarmin}
\ee
and yields for the ratio  $r=\tcarmin/\tcarmax$ 
\be
r = \frac{ \bip \Me }{3   a^2} 
\left[  1 + \left( \frac{6 b }{c^2} + 1 -\bip \right)  \frac{\bip}{a}  +  O\left( \frac{1}{a^2}\right) \right]. 
\label{eq:ratio}
\ee
Our main conclusion follows from those approximate expressions. 
 Large values of $\bce$ can induce, compared to the $\bce=0$ case, 
a very pronounced decrease in $\tcar$. 
For instance, $\bc=10^7$ yields $r=0.012$, that is a 100-fold reduction.

\begin{table}[htb]
  \begin{tabular}{r r r r r c c}
     \hline \hline
     $\bc$       & $\tcarmax$  & App. & FD   & $\tcarmin$  & App. & FD    \\   
     \hline
     $10^7$      &\ \ \ \ 10288 &    10314  &10100 &  \ \ \ \ \ \ 90  & 104 & 99      \\
  $5\ 10^7$      &    2213  &     2234  &   2020 &  88  & -   & 95        \\
     $10^8$      &    1208  &     1224  &   1010 &  86  & -   & 91        \\
     $10^9$      &     322  &      315  &    101 &  78  & -   & 50        \\
   \hline\hline
  \end{tabular}
\caption{Maximum and minimum decay times $\tcarmax$ and $\tcarmin$. 
App. and FD correspond to Eqs.~(\ref{eq:tcarmax}) and (\ref{eq:tcarmaxtoy}) for the former,  
and Eqs.~(\ref{eq:tcarmin}) and (\ref{eq:tcarmintoy}) for the latter. The conductance $\bc$ is expressed in W m$^{-2}$ K$^{-1}$ and the times in ps. }
\label{tab:tcarmaxmin}
\end{table}

\begin{figure}
\includegraphics[width=7cm]{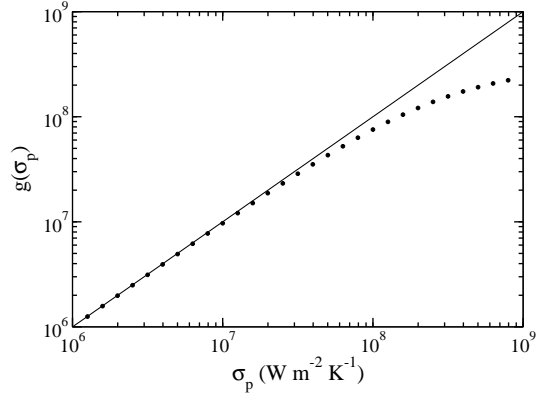}
\caption{The function  $g(\bc)$ as defined from  Eq.~\ref{eq:apptcar} (points). 
The line indicates the function $g(\bc)=\bc$. 
}
\label{fig:gbc}
\end{figure}

Table~\ref{tab:tcarmaxmin} compares exact numerical values for $\tcarmin$, $\tcarmax$, 
and their approximation given by Eqs.~\eqref{eq:tcarmax}-\eqref{eq:tcarmin}, 
which are surprisingly accurate, given the assumption on $\delta$ and $a$.  
Note that the approximate Eq.~\eqref{eq:tcarmin} predicts that 
$\tcarmin$ is independent of $\bc$. 
However, when taking the five-term expansion of $\tcarmin$,  
one recovers the slight decrease with $\bc$ which is observed for the exact $\tcarmin$. 
Finally, one can obtain a  good approximation
of $\tau$ with the  simple formula 
\be
\frac{\tcar - \tcarmin}{\tcarmax - \tcarmin } =  \frac{1}{1 + \frac{\bce}{ g(\bc)}}, 
\label{eq:apptcar}
\ee
where $ g(\bc)$ is a function plotted in  Fig.~\ref{fig:gbc}. 
For low $\bc\lesssim  5\times10^7$, $ g(\bc) \simeq \bc$ while it becomes smaller at higher $\bc$.

Two side remarks are in order. 
First, the formulas given above for the coefficients $a$, $b$ and $c$ 
are valid for any dimensionless parameters but may be simplified in many cases. 
For instance, taking the default  parameters of Tab.~\ref{tab:cases}, yields $\Me=1$, $\Mp=10$, $R=1000$, 
which suggests taking the limiting case $R \gg 1$, and $\Me \ll \Mp$,
and leads to the simple expressions
\be
a^2 &=& \Mp,  \qquad    b^2 = \frac{\Mp R^2}{4},  \qquad    c^2 = \Me R.  
\nonumber
\ee
Second, there is a simple limiting case for $\tcarmax$ as given by  Eq.~\ref{eq:tcarmax}. 
In addition to $R \gg 1$, 
let's assume that $\tcarmax \gg \taure$ and $\bip \ll 1$. 
With default parameters, the former approximation is well justified 
and the latter applies only for low $\bc$ values ($\bip=0.1$ for $\bc=10^7$ for instance). 
Then the first term inside the brackets of Eq.~\eqref{eq:tcarmax} dominates over the others, 
and one finds  $\tcarmax= \taucp$, 
i.e. one recovers the characteristic time for the one-temperature model in the limit $\bip \rightarrow 0$, 
as given in Appendix~A.

\subsection*{Appendix C: thermal conductivity for silicon and silica.} 
\label{sec:appnum}
\begin{figure}[h]
\includegraphics[scale=0.31]{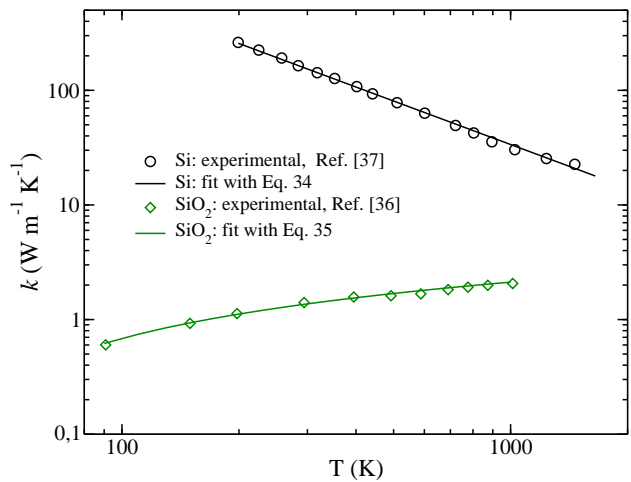}
\caption{(Color online) Silicon and silica bulk thermal conductivities extracted from the experimental measurements 
Refs.~\cite{Jund1999} and~\cite{Glassbrenner1964}.  
The solid lines are the fits obtained from Eqs.~\ref{eq_fit_ksilicon} and \ref{eq_fit_ksilica}.}
\label{fig_Thermal_conductivity}
\end{figure}

We have used the following fits to describe the temperature dependence of the thermal conductivity:
\begin{eqnarray}
k_{\rm Si}    &=& 203913 \  T^{-1.26},  \label{eq_fit_ksilicon} \\
k_{\rm SiO_2} &=& 0.624 \ln{T}-2.19,    \label{eq_fit_ksilica}
\label{eq_fit_k}
\end{eqnarray}
where the temperature $T$ is expressed in K and the conductivities in W\,m$^{-1}$\,K$^{-1}$.  
The experimental values and fitting curves are displayed in Fig.~\ref{fig_Thermal_conductivity}. 


\bibliography{bddloc.bib}

\end{document}